# Broadband diffraction of correlated photons from crystal superlattices

Zi S.D. Toa[1], Anna V. Paterova, Leonid A. Krivitsky

*Institute of Materials Research and Engineering (IMRE), Agency for Science, Technology and Research (A*STAR), Singapore 138634, Singapore*

Sources of broadband quantum correlated photons present a valuable resource for quantum metrology, sensing, and communication. Here, we report the generation of spectrally broadband correlated photons from frequency nondegenerate spontaneous parametric down-conversion in a custom-designed lithium niobate superlattice. The superlattice induces a nonlinear interference between the pump, signal and idler, resulting in an experimentally observed comb-like emission spanning 0.060 μm and 1.4 μm of spectral bandwidth at 0.647 μm and 3.0 μm wavelengths, respectively. While this broad mid-infrared bandwidth is attractive to quantum metrology and sensing due to the enablement of fast spectral multiplexing for data acquisition, the comb-like structure, achieved without an input frequency comb, offers targeted frequencies for quantum communication applications. In addition to useful technological applications, our concept offers an interesting analogy between optical diffraction in quantum and classical optics.

In classical optics, diffraction refers to phenomena in which a wave encounters an obstacle or aperture, which, in turn, effectively serves as a secondary source of waves (1). The amplitudes of these secondary waves subsequently interfere. In the case of light waves incident on an array of circular apertures, this results in a characteristic diffraction pattern comprising concentric rings of bright and dark regions that can be observed on the screen.

---

[1] Corresponding author: desmond_toa@imre.a-star.edu.sg

Here, we consider an analogous phenomenon in quantum optics, in which a coherent laser interacts with an array of nonlinear crystals, referred here as a superlattice. Each superlattice element analogously serves as a secondary source of waves, producing correlated photons via spontaneous parametric down-conversion (SPDC) (2-4). The wavefunctions of SPDC photons generated in the different elements interfere to produce an interference pattern in wavelength-angular ($\lambda$-k) coordinates (5). Furthermore, similar to conventional diffraction, interference fringe narrowing is observed with an increasing number of nonlinear elements. Thus, one can consider that the pump laser undergoes diffraction (in frequency coordinates) upon encountering the superlattice to give visible signal and mid-infrared (MIR) photons at distinct frequencies.

To demonstrate this effect, we custom-designed and fabricated a bi-periodically poled lithium niobate ($LiNbO_3$) (biPPLN) superlattice. The periodically poled regions, where the SPDC occurs, are separated by similarly poled regions of the $LiNbO_3$ crystal, see Figure 1. Thus, two periods, one for domains within stacks and the other for gaps, are operative here. Note that previous realizations of the superlattices had intervening linear media, which resulted in angular modulation of the $\lambda$-k spectrum (2). In contrast, the nonlinear interference in our biPPLN superlattice results in frequency modulated SPDC emission, which is highly relevant to applications in quantum key distribution, quantum metrology, and sensing. For example, utilizing spectral multiplexing, sensors can benefit from an increased measurement speed, improving their competitiveness with existing classical techniques. We also note here that biPPLNs have been used earlier to generate polarization entangled states (6, 7) and for optical wavelength conversion (8).

The theoretical study of the SPDC frequency modulation in crystal superlattices was presented in (5). The closest earlier experimental demonstration used two nonlinear fibers generating frequency degenerate photon pairs (9). However, the experimental realization of the superlattice with more nonlinear elements was limited by the stability and scalability of the fiber platform. In contrast, our approach of using a monolithic crystal enables easy scaling up of the number of nonlinear elements, with up to 86 elements demonstrated in this work. Furthermore, our approach possesses a high intrinsic stability, versatility of crystal design, and compatibility with $LiNbO_3$-based quantum integrated photonics (10-12). This platform enables us to experimentally demonstrate ultra-broadband frequency modulation with a signal full-width-half-maximum (FWHM) of 0.030 μm centered at 0.647 μm from a single crystal. This translates to a corresponding frequency modulation with FWHM of ~0.660 μm centered at 3.0 μm in the MIR idler channel, making it relevant to practical sensing and spectroscopic applications demanding ultra-broad bandwidth in the MIR, in particular those exploiting "undetected photons" (13-18).

The theory for SPDC in layered nonlinear media, such as the biPPLN superlattice here, was previously discussed (3-5, 19-22), but primarily limited to only two nonlinear crystals separated by a linear medium (3, 4). Nonetheless, here we briefly describe the one-dimensional theory.

The two-photon (or biphoton) state generated via SPDC in layers of nonlinear media is given by (22, 23)

$$|\psi\rangle = |vac\rangle + \sum_n \sum_{k_s,k_i} f_n(\vec{k_s}, \vec{k_i}) \hat{a}^+_{k_{n,s}} \hat{a}^+_{i_{n,s}} |vac\rangle, \qquad (1)$$

where $f_n(\vec{k_s}, \vec{k_i})$ is the biphoton field amplitude from the $n^{th}$ crystal, $\hat{a}^+_{k_{n,s}}$ and $\hat{a}^+_{i_{n,s}}$ are creation operators of the signal and idler photons in the $n^{th}$ crystal with wavevectors $\vec{k_s}$ and $\vec{k_i}$ respectively, and $|vac\rangle$ is the vacuum state (22, 24, 25).

The biPPLN superlattice is comprised of multiple nonlinear domains with the same material properties. We assume the pump to be a monochromatic plane wave. Thus, the amplitude of a biphoton field is given by $f_n \propto \chi_n E_p \int_{z_n}^{z_n+l_n} dz\, D^*_{n,p} D_{n,s} D_{n,i}$, where $\chi_n$ is the signed second-order optical susceptibility of the $n^{th}$ nonlinear domain, $z_n = \sum_{n'=1}^{n} l_{n'}$ and $z_n + l_n$ are, respectively, the front and back edge coordinates of the $n^{th}$ nonlinear crystal domain of thickness $l_n$, $E_p$ is the pump field (21, 22). $D_{n,\lambda}$ is the propagation function for the pump, signal, and idler ($\lambda = p, s, i$) wavelengths in the $n^{th}$ crystal domain as given by $D_{n,\lambda}(k_\lambda, z) = \exp(-ik_{n,\lambda}z)$, where $k_{n,\lambda}$ is the longitudinal wavevector of the wavelength $\lambda$ inside the $n^{th}$ nonlinear layer.

The two-photon intensity (normalized to the square of the crystal length) is thus given by the absolute square of the sum of contributions from the N crystal domains (3, 5, 22):

$$I(\omega, \theta) \propto \frac{1}{l^2} \left| \sum_{n=1}^{N} \chi_n l_n \, \text{sinc}(\Delta_n/2) \exp\left(-\frac{i\Delta_n}{2} + i\sum_{n'=1}^{n} \Delta_{n'}\right) \right|^2, \qquad (2)$$

where $l_n$ is the thickness of the $n^{th}$ crystal domain, $\Delta_n \equiv l_n(-k_{p,z} + k_{s,z} + k_{i,z})$ is the longitudinal phase mismatch in each domain, and $l = \sum_{n=1}^{N} l_n$ is the total thickness of the crystal. Note that the SPDC intensity thus normalized gives us a good way to compare down-conversion efficiencies of different designs. Note that equation (2) also implies that the interference between the correlated signal and idler photons results in the experimentally observed signal frequency modulation.

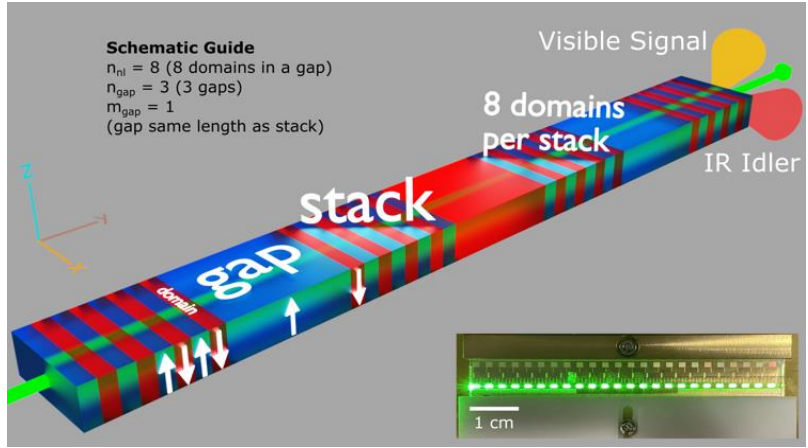

Figure 1. Structure of a biPPLN superlattice with eight domains per stack ($n_{nl} = 8,$), three gaps ($n_{gap} = 3$) of the same length as a stack ($m_{gap} = 1$). The white arrows denote the optic axis orientation in each domain. Blue and red colored domains have up and down optic axis orientation, respectively. Green pump laser propagates in the positive $y$ direction. Note that the LiNbO$_3$ material fills the gap. (Inset) Photograph of the LiNbO$_3$ crystal with one poling design illuminated by the pump laser. The bright scattering regions indicate different stacks.

The length of each domain in the poled region of the LiNbO$_3$ crystal is $l_{domain} = 5.16\ \mu m$, chosen to fulfill the SPDC phase-matching conditions. Upon pumping with a 0.532 μm continuous wave laser, the structure yields frequency nondegenerate SPDC centered at 0.647 μm and 3.0 μm for the signal and idler photons, respectively.

The domains are arranged into stacks, such that each stack contains $n_{nl}$ domains with alternating optic axes between adjacent domains, see Figure 1. Thus, the length of each stack is $l_{stack} = n_{nl} \times l_{domain}$. In between two stacks, the intervening gap is specified with a length $m_{gap}$ times that of the stack and has an optical axis orientation opposite to the adjacent domains of two different stacks. Thus, the length of each gap is $l_{gap} = m_{gap} \times l_{stack}$. Note that stacks with an even number of domains will result in alternating gap optic axis orientations. Each poling design in the crystal is comprised of $n_{gap}$ gaps and $n_{stack} = (n_{gap} + 1)$ stacks. The resultant length of a poling design is thus given by

$$l_{design} = l_{stack} \times (n_{gap} \times m_{gap} + n_{gap} + 1), \tag{3}$$

Thus, each poling design can be uniquely specified using the parameters $n_{nl}$, $n_{gap}$ and $m_{gap}$, with a fixed $l_{domain}$.

The specifications for the various poling designs are summarized in Table 1.

A 63.5 mm × 8.2 mm × 0.5 mm (length x width x height) LiNbO₃ crystal was designed by us, and custom manufactured by HC Photonics. Three poling designs are presented here, see Table 1. The crystal facets are coated with an anti-reflective (AR) coating with less than 0.5% reflectance for 0.532 µm, 0.647 µm, and 3.0 µm wavelengths.

Table 1. Parameters and experimental results for three poling designs. Agreement between the theoretical model and experimental results is described by a Sample Pearson correlation coefficient (SPCC).

| Design | Design Parameters | | | Experimental Results | | | | |
|---|---|---|---|---|---|---|---|---|
| | $n_{nl}$ | $n_{gap}$ | $m_{gap}$ | Signal envelope FWHM (µm) | Idler envelope FWHM (µm) | Signal average peak spacing (µm) | Idler average peak spacing (µm) | SPCC |
| 1 (Fig. 2a) | 16 | 85 | 8 | 0.0304 | 0.657 | 0.0041 | 0.090 | 0.920 |
| 2 (Fig. 2b) | 64 | 21 | 8 | 0.0081 | 0.173 | 0.0010 | 0.022 | 0.877 |
| 3 (Fig. 2c) | 16 | 23 | 32 | 0.0307 | 0.666 | 0.0011 | 0.025 | 0.838 |

In our experiments, the crystal is pumped by a 0.532 µm wavelength single longitudinal mode (SLM) continuous-wave Nd:YAG laser (200 mW, linewidth ≤ 1MHz, Oxxius SA). A 2× beam expander is placed before the crystal to reduce the transverse dimension of the pump beam to account for the finite thickness of the crystal. A fraction of pump photons is down-converted to ~0.647 µm visible signal and ~3.0 µm MIR idler photons via type-0 collinear frequency nondegenerate SPDC. A notch filter is used to block the pump beam after the crystal, while the signal photons are focused onto the slit of an imaging spectrograph (Acton SpectraPro 2300i) using a three-lens system. The spectrum of signal photons is centered at 0.647 μm and is recorded using a visible-range electron-multiplying charge-coupled device (EMCCD) camera (Andor iXon 897) at the output of the spectrograph. Camera exposure duration varies from 5 s to 10 s, with a 100× electron-multiplying gain set during the data acquisition. The EMCCD sensor has 512 × 512 pixels with pixel size of 16 μm. and was kept at a temperature of -80°C. All measurements were conducted at the ambient lab temperature of 22°C.

Superlattice spectra simulation code is written in Python 3.7 (26, 27) with refractive indices from Gayer, *et al.* (28) set to the ambient lab temperature of 22°C. Furthermore, Gaussian convolution is employed in our spectra, in order to match detection of the 0.532 μm SLM pump laser beam over 5×5 pixels of the EMCCD camera.

We observed the ultra-broadband frequency comb-like diffraction, spanning 0.0304 μm and 0.0307 μm FWHM in signal wavelength, from designs **1** and **3**, see Figure 2a, c. The overall spectral bandwidth, defined by the 'envelope' formed by the maxima of the spectral fringes (green line in Figure 2), was similar between these two designs as they share the same number of nonlinear domains in a stack. This envelope FWHM is determined from fitting a Gaussian curve to the experimental fringe peaks, see Figure 2a, c. The observed difference in the modulation period is due to different gap lengths between stacks. This has a clear analogy with classical diffraction, in which closer spaced apertures result in the diffraction pattern with less frequent modulation.

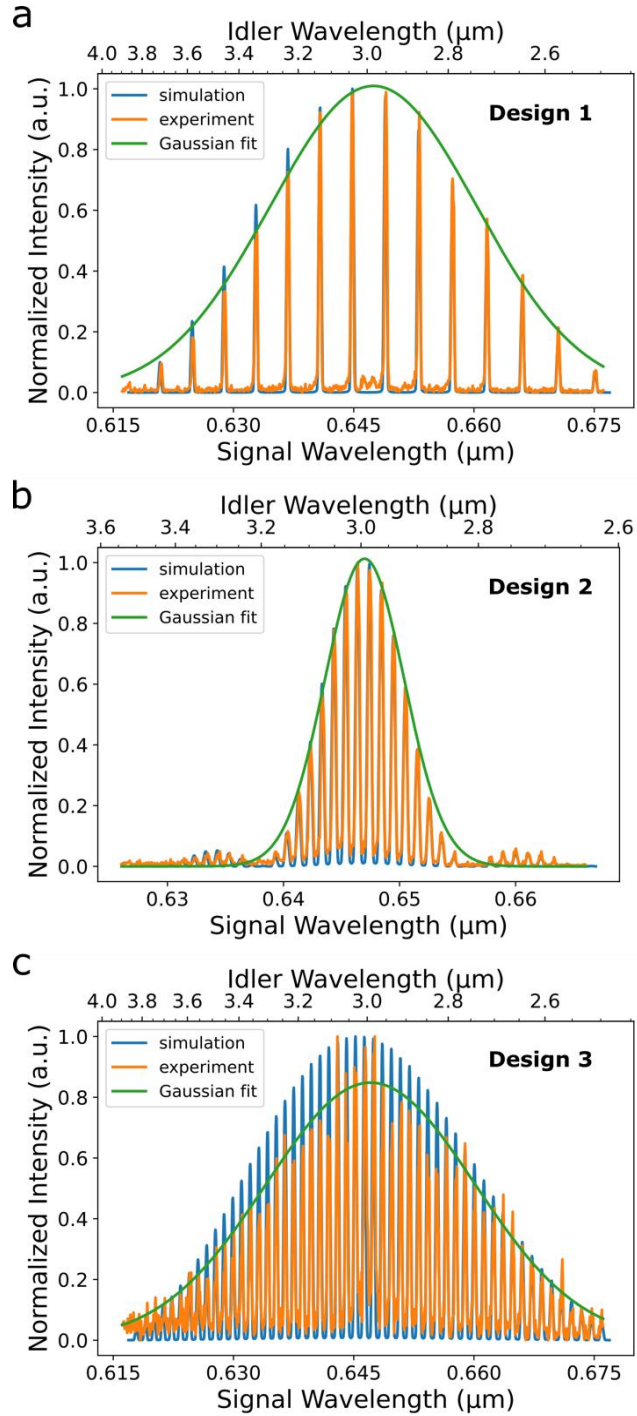

Figure 2. Normalized experimental (orange line) and simulated (blue line) spectral cross-sections at zero scattering angle for designs (a) 1, (b) 2, and (c) 3, see Table 1. The green line is a Gaussian fit to the maxima in the experimental data. Note the ultra-broadband diffraction from designs **1** and **3** and the dependence of the spectral envelope on the parameters of the superlattice. Sub-figure (b) is presented in a different horizontal scale for clarity.

The fact that the number of nonlinear domains per stack determines the overall spectral bandwidth was further demonstrated with design **2**, see Figure 2b. A larger number of domains per stack led to shrinkage of the overall spectral bandwidth, despite having a similar number of gaps with design **3** and the same gap length multiple as design **1**, see Figure 2a, c. Thus, broader spectral bandwidth than presented here is achievable with fewer nonlinear domains per stack. Again, these observations are akin to classical diffraction, where gratings with fewer elements produce narrower diffraction patterns. We emphasize that the observed diffraction patterns arise from the quantum interference of the biphoton wavefunction, see equation (2).

Agreement between observed and simulated spectra was generally good, as evidenced by the match between spectral positions of the comb peaks and the high Sample Pearson Correlation Coefficients (SPCC) for the three designs, see Table 1. In Figure S1 of Supplemental Material, we show the full λ-k spectra produced by the superlattices. The spectra indicate additional angular modulation at larger scattering angles (more than 1° at fixed wavelengths, see also Figure S2. These results indicate the richness of the proposed platform for spectral and spatial shaping of the biphoton emission.

In summary, we have thus demonstrated ultra-broadband diffraction in frequency coordinates from the nonlinear interference of SPDC photons. For this, we custom-designed and fabricated biPPLN superlattices, with which SPDC emission spanning 0.060 μm and 1.4 μm of spectral bandwidth around 0.647 μm and 3.0 μm wavelengths, respectively, was demonstrated. This is competitive with recent work by Vanselow *et al.* in which similar bandwidths in the visible and MIR were demonstrated (29). Conversely, our superlattice design is not limited to specific pump wavelengths arising from requiring matching group indices. In addition, our superlattice design can achieve greater bandwidth by simply decreasing the number of domains per stack. Note that the achievable wavelength range can be further increased via temperature-tuning, see Figure S3 in Supplemental Material. Beyond the demonstration of an interesting analogy between quantum and classical optical diffraction, which is of fundamental interest, these biPPLN superlattices have practical implications. They include but are not limited to emerging applications requiring broad spectral bandwidth (29-34), such as broadband IR sensing (2), metrology (35), and quantum entangled frequency combs (36, 37).

For example, our light source can be embedded in a nonlinear interferometer in which mid-IR spectral fingerprints can be readout from the observation of correlated visible range signal photons (35). The crystal superlattice will provide multiple spectral components in the mid-IR, which could then be read out at once, thus eliminating the need for time consuming spectral scanning. It has not escaped our notice that the designs presented herein might make for easier targeted frequency multiplexing for quantum communication and quantum key distribution (38). Last but not least, our approach, akin to quantum mode engineering (39, 40), can potentially improve other instrument specifications in quantum hyperspectral IR imaging (13, 14, 16-18).

The authors acknowledge the support of the Agency for Science, Technology and Research (A*STAR) under its project C210917001.. This work was also supported by the A*STAR Computational Resource Centre through the use of its high-performance computing facilities.

# Supplemental Material:

# Broadband diffraction of correlated photons from crystal superlattices

Zi S.D. Toa, Anna V. Paterova, Leonid A. Krivitsky

*Institute of Materials Research and Engineering (IMRE), Agency for Science, Technology and Research (A\*STAR), Singapore 138634, Singapore*

**Experimental wavelength-angular spectra**

Spontaneous parametric down-conversion (SPDC) is not limited to the zero scattering angle (i.e. in collinear geometry), as phase-matching conditions are also fulfilled between pump photons and transversely scattered signal and idler photons at non-zero scattering angles. Here in Figure S3, we present SPDC emission over wavelength and external scattering angles in the signal channel, showing the agreement between simulations and experimental observations, which are discussed in details in the main text

In our case the gaps between the nonlinear elements of the superlattice are filled by the same material. This results in pronounced modulation in frequency coordinate for small scattering angles (less than ±1°). At larger scattering angles (over ±1°), we can observe an angular modulation at a given wavelength due to the curvature of the SPDC spectral curves. We can resolve this modulation in the range of angles between ±1° and ±2°, see Figure S4. At larger scattering angles, the resolution of the experimental imaging system is not enough to resolve the fine angular modulation, see Figure S3 d, f and Figure S4. These results indicate the richness of the proposed platform for spectral shaping of the SPDC emission.

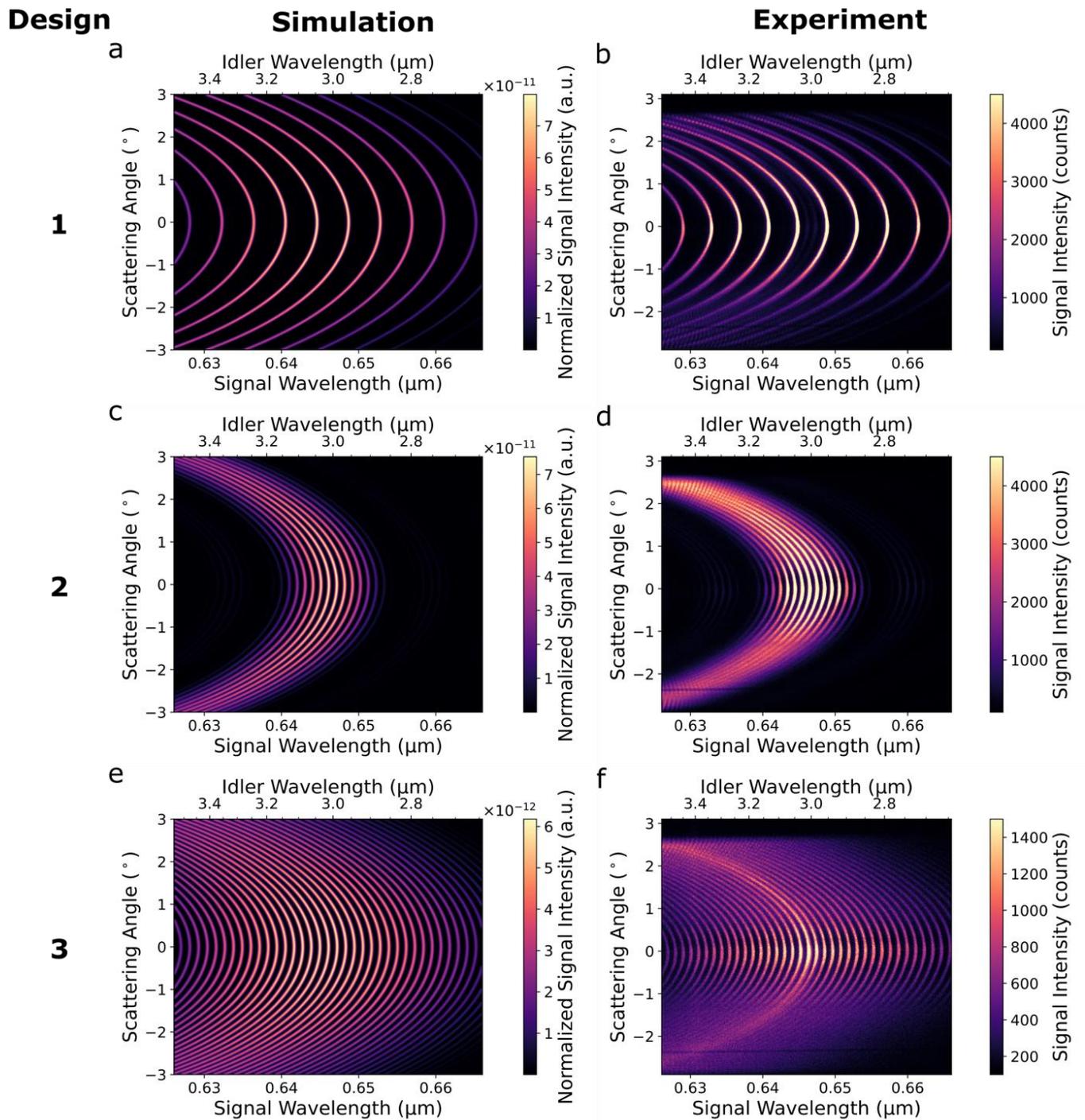

Figure S3. Simulated (a, c, e) and (b, d, f) experimental wavelength-angular spectra of SPDC from designs (a, b) 1, (c, d) 2, and (e, f) 3. Note the good agreement between simulated and experimental spectra, particularly in scattering angles close to zero.

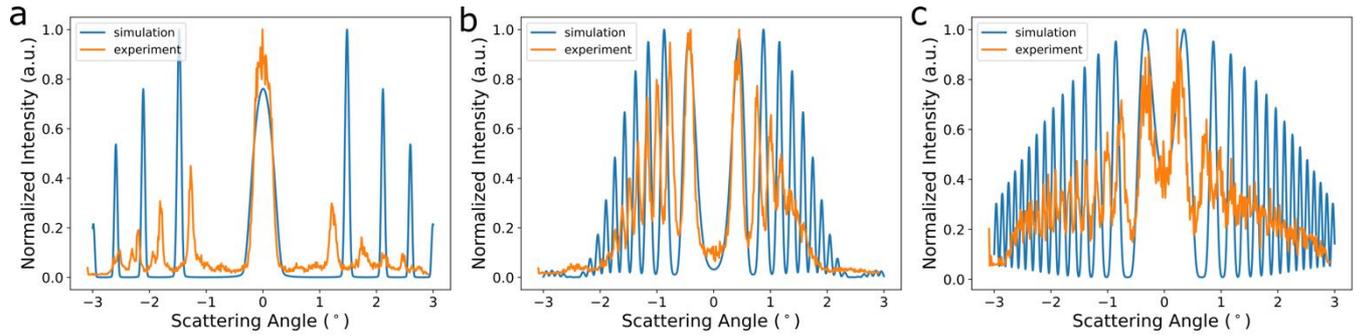

Figure S4. Normalized experimental (orange line) and simulated (blue line) angular cross-sections at signal wavelength of 0.645 μm for designs (a) 1, (b) 2, and (c) 3. Note the modulation along scattering angles.

**Temperature-dependent tunability**

The wavelengths of the signal and idler from SPDC can be further adjusted via the use of an oven, as the SPDC phase-matching condition rests upon the temperature-dependent refractive indices of lithium niobate. Here in Figure S5, we demonstrate this temperature tunability, which increases the range of achievable wavelengths on top of the broad spectral SPDC from the superlattice design.

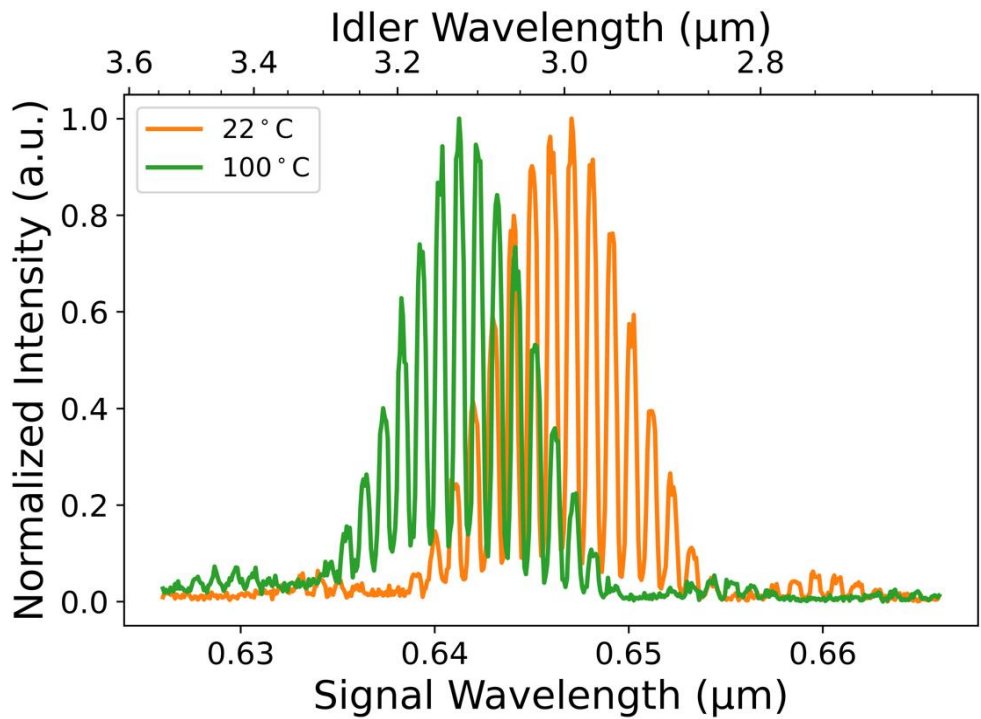

Figure S5. Normalized experimental spectral cross-sections at zero scattering angle from design 2 at 22°C (orange line) and 100°C (green line). Note how this temperature difference resulted in a spectral difference of ~0.005 μm and ~0.2 μm in the signal and idler channels respectively.